\documentclass{achemso}

\usepackage{bm,mathrsfs,dcolumn,graphicx,color}
\usepackage{amsmath}
\usepackage{graphicx}
\usepackage[T1]{fontenc}
\usepackage{hyphenat}
\usepackage{setspace}
\usepackage[normalem]{ulem}
\usepackage[colorlinks=true,citecolor=darkerblue,linkcolor=darkerred]{hyperref}
\usepackage{hyperref}
\usepackage{enumitem}

\definecolor{ablue}{rgb}{0.1,0.35,0.75}

\definecolor{dgreen}{rgb}{0,0.65,0.2}

\definecolor{darkerblue}{rgb}{0,0,0.75}
\definecolor{darkerred}{rgb}{0.8,0,0}
\usepackage[colorlinks=true,citecolor=darkerblue,linkcolor=darkerred]{hyperref}

\usepackage[toc,page]{appendix}
\newcommand{\nocontentsline}[3]{}
\newcommand{\tocless}[2]{\bgroup\let\addcontentsline=\nocontentsline#1{#2}\egroup}

\newcommand{\fig}[1]{Fig.\,S\ref{#1}}

\makeatletter
\renewcommand{\p@subsection}{}
\renewcommand{\p@subsubsection}{}
\makeatother

\author{Jonas D. Ziegler}
\author{Jonas Zipfel}
\author{Barbara Meisinger}
\affiliation{Department of Physics, University of Regensburg, Regensburg D-93053, Germany}
\author{Matan Menahem}
\affiliation{Department of Materials and Interfaces, Weizmann Institute of Science, Rehovot, Israel}
\author{Xiangzhou Zhu}
\affiliation{Department of Physics, Technical University of Munich, 85748 Garching, Germany}
\author{Takashi Taniguchi}
\affiliation{International Center for Materials Nanoarchitectonics,  National Institute for Materials Science, Tsukuba, Ibaraki 305-004, Japan}
\author{Kenji Watanabe}
\affiliation{Research Center for Functional Materials, National Institute for Materials Science, Tsukuba, Ibaraki 305-004, Japan}
\author{Omer Yaffe}
\affiliation{Department of Materials and Interfaces, Weizmann Institute of Science, Rehovot, Israel}
\author{David A. Egger}
\affiliation{Department of Physics, Technical University of Munich, 85748 Garching, Germany}
\author{Alexey Chernikov}
\email{alexey.chernikov@ur.de}
\affiliation{Department of Physics, University of Regensburg, Regensburg D-93053, Germany}

\title{Fast and Anomalous Exciton Diffusion in Two-dimensional Hybrid Perovskites}

\begin{document}

\begin{abstract}
Two-dimensional hybrid perovskites are currently in the spotlight of condensed matter and nanotechnology research due to their intriguing optoelectronic and vibrational properties with emerging potential for light-harvesting and -emitting applications. 
While it is known that these natural quantum wells host tightly bound excitons, the mobilities of these fundamental optical excitations at the heart of the optoelectronic applications are still largely unexplored.
Here, we directly monitor the diffusion of excitons through ultrafast emission microscopy from liquid helium to room temperature in hBN-encapsulated two-dimensional hybrid perovskites. 
We find very fast diffusion with characteristic hallmarks of free exciton propagation for all temperatures above 50\,K.
In the cryogenic regime we observe nonlinear, anomalous behavior with an exceptionally rapid expansion of the exciton cloud followed by a very slow and even negative effective diffusion.
We discuss our findings in view of efficient exciton-phonon coupling, highlighting two-dimensional hybrids as promising platforms for many-body physics research and optoelectronic applications on the nanoscale.
\\
\\
\end{abstract}
\maketitle

Nanostructured, two-dimensional (2D) organic-inorganic perovskites, studied as early as in the 1980's and 90's\,\cite{Ishihara1989, Mitzi1994, Muljarov1995, Kagan1999}, have recently emerged as promising candidates for high-efficiency photovoltaic and light-emitting applications\,\cite{Kumar2016,Tsai2016,Wang2016b,Tsai2018,Chen2018,Xing2018}.
The renewed attention is strongly motivated by the structural and chemical design flexibility of these natural multi-quantum-well systems\,\cite{Mitzi1999}, promising to accelerate the ongoing search for improved device performance and stability\,\cite{Zhang2019c,OrtizCervantes2019}.
Of key interest to this challenge are intriguing phenomena that revolve around coupling between electronic and vibronic excitations\,\cite{Ishihara1989,Tanaka2005,Gauthron2010,Yaffe2015,Guo2016,Wright2016,Ni2017,Straus2018,Blancon2018,Baranowski2019,Cho2019,SrimathKandada2020}.
The 2D confinement of the electronic states in the inorganic layers together with reduced dielectric screening from the organic barriers give rise to an exceptionally strong Coulomb interaction and tightly bound exciton states with binding energies on the order of many 100's of meV\,\cite{Ishihara1989,Tanaka2005,Yaffe2015,Blancon2018}.
In addition, similar to their three-dimensional (3D) analogues\,\cite{Egger2018}, the vibronic characteristics of the 2D perovskites are peculiar in contrast to conventional inorganic semiconductors, since they are unusually soft.
As a consequence, the coupling of the excitonic quasiparticles to lattice vibrations is very efficient, resulting in fast exciton-phonon scattering\,\cite{Straus2016,Wright2016} and the proposed emergence of polaron effects\,\cite{Gauthron2010,Guo2016,SrimathKandada2020}.
This combination renders 2D perovskites a highly interesting material platform to study fundamental many-particle phenomena in hybrid systems motivated by both basic science and emerging technologies.

A key process directly influenced by these properties is the diffusion of optical excitations that is intimately tied to the fundamental understanding of the exciton structure, spatial energy landscape, and exciton-phonon interactions.
As a consequence, it is also one of the most crucial parameters ultimately determining efficiency limits of optoelectronic devices, including both light-harvesting and -emitting applications.
Interestingly, in contrast to 3D perovskites\,\cite{Delor2020, Baranowski2020}, the exciton propagation remains barely explored for 2D variants, with the initial reports of room-temperature diffusion\,\cite{Deng2020,Seitz2020} revealing the role of the organic compounds\,\cite{Seitz2020} and exciton-exciton interactions\,\cite{Deng2020}.
The nature of the exciton transport, however, remains only little understood with major open questions regarding conditions for localization and free propagation as well as fundamental limits for the exciton diffusivity.

\begin{figure*}[t]
	\centering
			\includegraphics[width=16 cm]{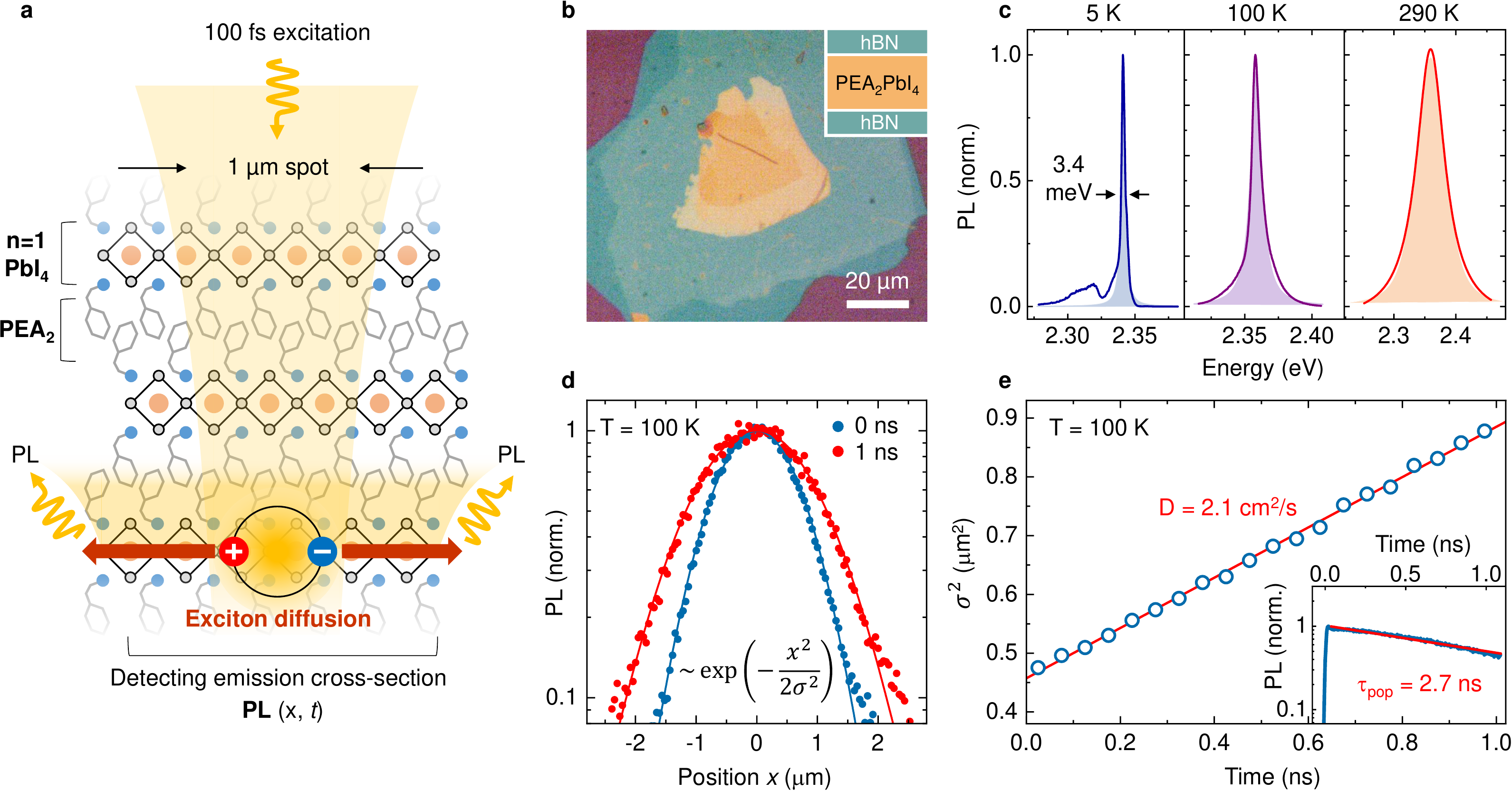}
		\caption{Exciton diffusion microscopy of 2D perovskites.
		a. Schematic illustration of the exciton diffusion in 2D ($n$\,=\,1) phenylethylammonium-lead-iodide (PEA$_2$PbI$_4$) studied by spatially- and time-resolved photoluminescence microscopy. 
		b. Optical micrograph of a typical, 50\,-\,80\,nm thin sample encapsulated between two layers of hBN. 
		c. Selected continuous-wave PL spectra recorded at temperatures of 5, 100, and 290\,K, presented together with Lorentz peak profiles.
		d. Representative spatial profiles of the sample PL for 0 and 1\,ns after pulsed excitation at $T=100$\,K for the excitation density of 2.6\,nJ/cm$^2$ per pulse. 
		e. Corresponding time-dependent variance $\sigma^2(t)$ of the PL profile as function of time; transient intensity is shown in the inset. 
		Extracted diffusivity of 2.1\,cm$^2$/s and population lifetime of 2.7\,ns obtained from the PL transient (corrected for diffusion of excitons out of the detection area) are indicated.
		}
	\label{fig1}
\end{figure*} 
Here, we directly monitor exciton propagation in thin sheets of 2D perovskites via time- and spatially-resolved emission microscopy across the temperature range from 5 to 290\,K.
Already at room temperature we consistently observe very efficient exciton transport with the diffusivity of 1\,cm$^2$/s, corresponding to an effective exciton mobility of 40\,cm$^2$/Vs.
Lowering of temperature leads to a strong increase of the propagation efficiency for all temperatures down to 50\,K, characteristic for \textit{free exciton} transport, with the diffusivity and effective mobility reaching values as high as 3\,cm$^2$/s and 1000\,cm$^2$/Vs, respectively.
Approaching 5\,K, the measurements reveal an intriguing non-linear regime of \textit{anomalous} diffusion showing a rapidly expanding exciton cloud and diffusivities up to 30\,cm$^2$/s.
It is followed by an effective shrinking of the spatial distribution corresponding to \textit{negative} diffusion.
The subsequent suppression of the exciton propagation for the remaining, long-lived exciton fraction exhibits hallmarks of localization and thermal activation.
Combining optical transport data with scattering rate analysis from spectral broadening and exciton mass from first-principles calculations we estimate the diffusivity limits for the free exciton propagation in a semi-classical model, rationalizing experimental findings.
It allows us to propose a broader framework for the fundamental understanding of the optical transport in 2D perovskites.
 
\textbf{Ultrafast Exciton Microscopy.}
To study exciton propagation we employ the technique of time- and spatially-resolved emission microscopy\,\cite{Akselrod2014,Ginsberg2020}, schematically illustrated in Fig.\,\ref{fig1}a.
The excitons are optically injected by 100\,fs short laser pulses at 3.1\,eV photon energy and 80\,MHz repetition rate, focused on an area of about 1\,$\mu$m diameter. 
Time-resolved expansion of the exciton cloud is then recorded by imaging the luminescence cross-section onto a streak camera detector.
The layered perovskites under study are monocrystalline 2D phenylethylammonium-lead-iodide (PEA$_2$PbI$_4$) thin sheets that are considered for their lack of a temperature-induced phase transition\,\cite{Lekina2019}.
The flakes are chemically synthesized and mechanically exfoliated to thin layers with an average thickness in the range of 50 - 100\,nm, as estimated from the optical contrast and confirmed by atomic-force microscopy (see SI).
Using the polymer stamping technique initially developed for inorganic van der Waals materials\,\cite{Castellanos-Gomez2014} we fully encapsulate the 2D perovskites between thin sheets of hBN for environmental protection and improved stability under excitation\,\cite{Seitz2019}.
An optical micrograph of one of the studied samples is presented in Fig.\,\ref{fig1}b.
All experiments are performed in a microscopy cryostat under high vacuum conditions.
Further details are found in the Supplementary Information and Ref.\,\cite{Kulig2018}.

Representative emission spectra of hBN-encapsulated PEA$_2$PbI$_4$ are presented in Fig.\,\ref{fig1}c at low (5\,K), intermediate (100\,K) and high (290\,K) temperatures.
The emission is always dominated by a single exciton resonance at about 2.35\,eV, close to the main peak in absorption-type reflectance contrast measurements (see Supplementary Information (SI)), typical for the studied systems\,\cite{Ishihara1989}.
Only at lowest temperatures we detect weak sideband emission at lower energies indicating additional contributions from localized states or phonon-assisted processes\,\cite{Straus2016}.
Here we note that the encapsulation in hBN should not affect the material response too strongly due to the relatively large thickness of the studied samples resulting in the majority of the excitations being far from the interface.
We emphasize, however, that we consistently observe very narrow resonances both in emission and reflectance spectra at cryogenic temperatures with linewidths that are typically under 10\,meV and can be as low as 3\,meV, as shown in Fig.\,\ref{fig1}c.
This improvement of optical quality may indeed be related to a beneficial effect of the encapsulation due to suppression of degradation and, potentially, even static disorder in analogy to inorganic monolayers\,\cite{Raja2019}.

Typical spatial profiles of the exciton emission at $T=100$\,K are presented in Fig.\,\ref{fig1}d, as obtained from the streak camera images at 0 and 1\,ns time delay after the excitation.
The data are normalized for direct comparison and shows a pronounced, time-dependent broadening of the exciton in-plane distribution.
For each time step we fit the measured PL intensity profile with a Gauss function, $I_{PL}(x,t) \propto exp[-x^2/2\sigma(t)^2]$.
Here, $x$ is the spatial, one-dimensional coordinate along the direction of the detected emission cross-section.
The extracted variance $\sigma(t)^2$ is presented in Fig.\,\ref{fig1}e and increases linearly with time, characteristic for conventional diffusion\,\cite{Ginsberg2020}.

The diffusion coefficient $D=2.1$\,cm$^2$/s is extracted from the slope of the variance measured along the one-dimensional coordinate according to $\Delta\sigma^2=\sigma^2(t)-\sigma^2(0)=2Dt$\,\cite{Ginsberg2020}.
The quantity $\Delta\sigma^2$ is commonly known as mean squared displacement (MSD) that accounts for the time-dependent component of the spatial broadening.
While the excitons are neutral quasiparticles it is still useful to express the diffusivity also in terms of effective mobility $\mu$ for direct comparison with the electron transport.
From the Einstein relation\,\cite{Ashcroft1976} $\mu=D\times e/k_BT$, where $k_B$ is the Boltzmann constant and $e$ the elementary charge we obtain the effective exciton mobility at $T=100$\,K of 250\,cm$^2$/Vs.
The corresponding diffusion length is 1.06\,$\mu$m, defined as $\sqrt{2D\tau_{pop}}$\,\cite{Ginsberg2020}. 
Here we use the population lifetime $\tau_{pop}=2.7$\,ns extracted from the PL transient taking into account an additional decay channel due to exciton diffusion out of the imaging area (see SI).
In the following the technique of the spatially- and time-resolved microscopy is employed for the investigation of exciton propagation across linear and nonlinear regimes.

\begin{figure}[t]
	\centering
			\includegraphics[width=10 cm]{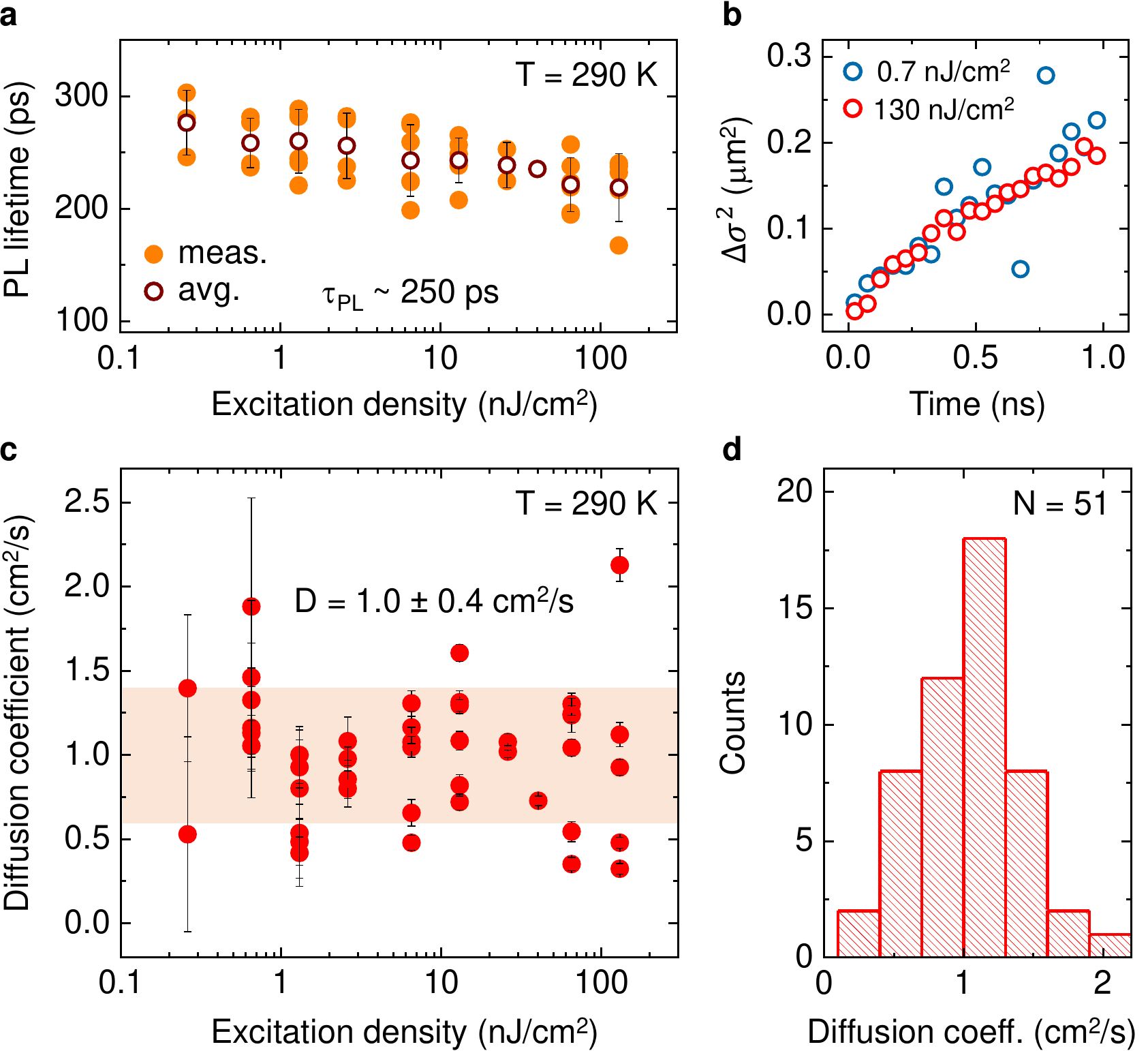}
		\caption{Room temperature exciton diffusion.  
		a. PL lifetimes as function of excitation density per pulse. 
		b. Mean squared displacement $\Delta\sigma^2=\sigma^2(t)-\sigma^2(0)$ of the spatial profile for two different excitation densities. 
		a. Corresponding diffusion coefficients, measured on several flakes and sample positions. Shaded area indicates the average and the standard deviation within the studied sample set. 
		d. Histogram of all measured room temperature diffusion coefficients.
		}
	\label{fig2}
\end{figure}

\textbf{Room Temperature Exciton Diffusion.}
First, we study room temperature diffusion and investigate density-dependent dynamics of the exciton population, since exciton-exciton interactions are known to strongly affect measured effective diffusivity\,\cite{Warren2000,Kulig2018,Deng2020}.
Characteristic PL decay times are presented in Fig.\,\ref{fig2}a as a function of energy density per pulse between 0.3 and 130\,nJ/cm$^2$ from individual measurements together with the arithmetic average.
The emission lifetime of about 250 ps\,remains almost constant across more than two orders of injection densities.
Given the weak excitation conditions used in our pulsed experiments this observation is in agreement with the exciton-exciton annihilation coefficients recently reported for a similar material system, butylammonium lead-iodide\,\cite{Deng2020}.

Similarly, the exciton diffusion remains virtually unaffected when changing the excitation density as illustrated in Fig.\,\ref{fig2}b by two representative MSD traces recorded at different injection densities.
The complete data set of the measured diffusion constants, collected on different flakes and sample positions is presented in Fig.\,\ref{fig2}c and summarized in a histogram in Fig.\,\ref{fig2}d.
The extracted mean diffusivity is $1\pm0.4$\,cm$^2$/s (from the total number of counts $N=51$); the deviation is largely associated with variations between different sample positions.
It corresponds to an effective exciton mobility of $\mu=40$\,cm$^2$/Vs and a mean diffusion length on the order of 200\,nm.

For a hybrid, excitonic material at room temperature the exciton propagation is very fast with the diffusivity being several orders of magnitude larger compared to molecular crystals\,\cite{Akselrod2014,Delor2020}.
It is closer to the exciton diffusion found in inorganic van der Waals monolayer semiconductors\,\cite{Kumar2014,Yuan2017,Zipfel2019a} and is very similar to that of the free carriers in 3D perovskites\,\cite{Egger2018,Delor2020}.
Interestingly, the determined value for $\mu$ is above the effective charge-carrier mobility in the same system determined at higher pump densities\,\cite{Milot2016}.
We also note that while the absolute diffusivity in the measured samples is about an order of magnitude higher than recently observed for 2D perovskites\,\cite{Seitz2020,Deng2020} it is generally consistent with the trends presented in these reports.
In particular, as shown in Ref.\,\cite{Seitz2020}, 2D PEA$_2$PbI$_4$ crystals studied here exhibit higher exciton mobility due to increased stiffness of the lattice reducing the exciton-phonon coupling compared to other organic spacers.
In addition, the encapsulation in hBN may in principle further contribute to mechanical stiffness and, more importantly, influence and suppress static disorder in agreement with very narrow spectral linewidths observed in our samples at low temperatures. 
Finally, as we discuss in the analysis section further below, it appears reasonable that the fast room temperature diffusion on the order of 1\,cm$^2$/s is already largely determined by the exciton-phonon interaction.
It is thus likely to represent a fundamental upper limit for the room temperature diffusivity in the studied 2D perovskite.

\begin{figure*}[t]
	\centering
			\includegraphics[width=16.5 cm]{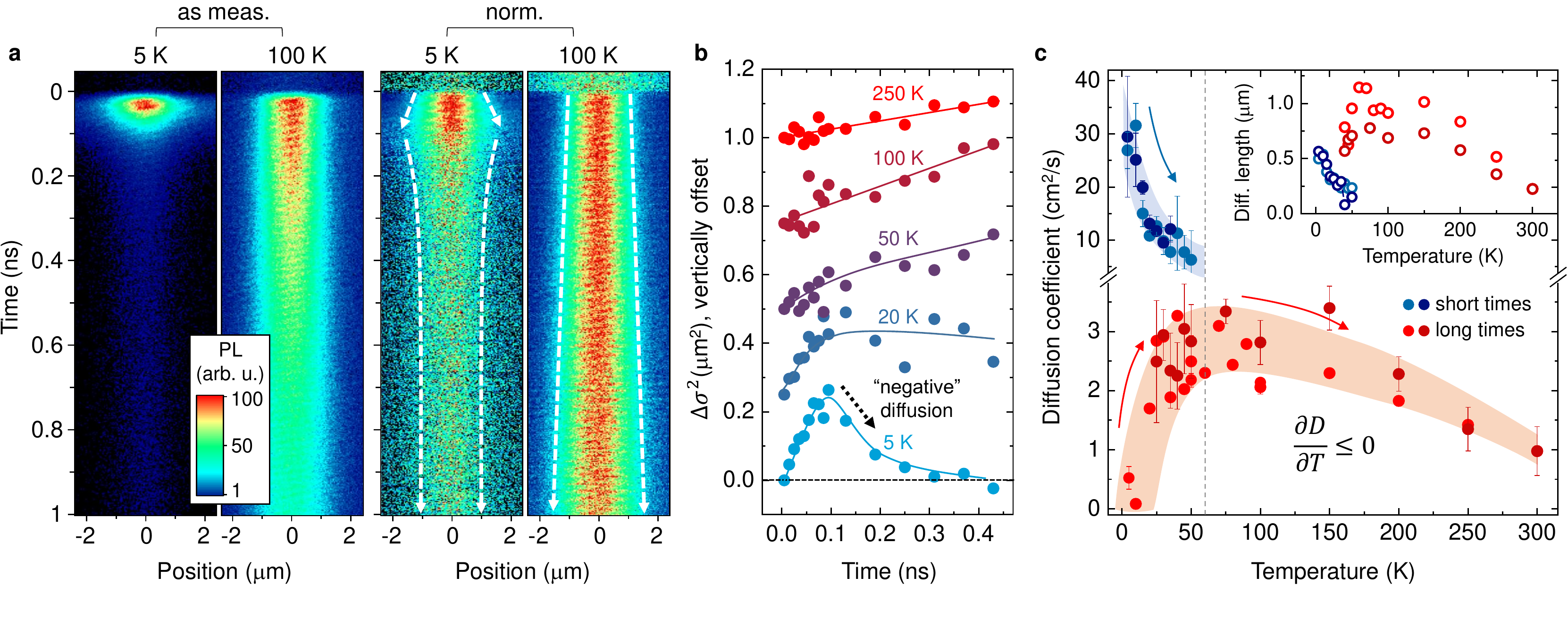}
		\caption{Temperature-dependent exciton propagation.
		a. Representative streak camera images of the spatially- and time-resolved PL for low (5\,K) and elevated (100\,K) temperature regimes for the excitation density of 2.6\,nJ/cm$^2$ per pulse. 
		The data is presented as-measured (left two panels) and normalized at each time step (right two panels).
		b. Time-resolved mean squared displacement $\Delta\sigma^2=\sigma^2(t)-\sigma^2(0)$ of the PL profiles for several selected temperatures between 5 and 250\,K. 
		c. Temperature-dependent diffusion coefficients at short (<50\,ps, blue dots) and very long times ($\gg$100\,ps, beyond ``negative'' diffusion regime, red dots) after the excitation. 
		Lighter and darker colors indicate data from two different samples.
		The inset shows the corresponding diffusion length $L$ obtained from measured diffusivity $D$ and PL lifetime 𝜏 according to $L=\sqrt{2D\tau_{pop}}$.
		Dashed and solid lines as well as the shaded areas in a, b, and c are guides to the eye.  
		}
	\label{fig3}
\end{figure*}

\textbf{Conventional and Anomalous Diffusion Regimes.}
Distinct regimes of exciton transport are revealed by lowering the sample temperature.
Characteristic, qualitative differences are illustrated by representative streak camera images showing spatially- (x-axis) and time-resolved (y-axis) emission presented in Fig.\,\ref{fig3}a for the temperatures of 5 and 100\,K in false-color plots.
As-measured luminescence counts (left panels) and data normalized at each time-step (right panels) emphasize population decay and spatial dynamics, respectively.
MSD traces for selected temperatures are shown in Fig.\,\ref{fig3}b, corresponding to the time-dependent changes of the spatial broadening of the PL.

\begin{figure}[t]
	\centering
			\includegraphics[width=11 cm]{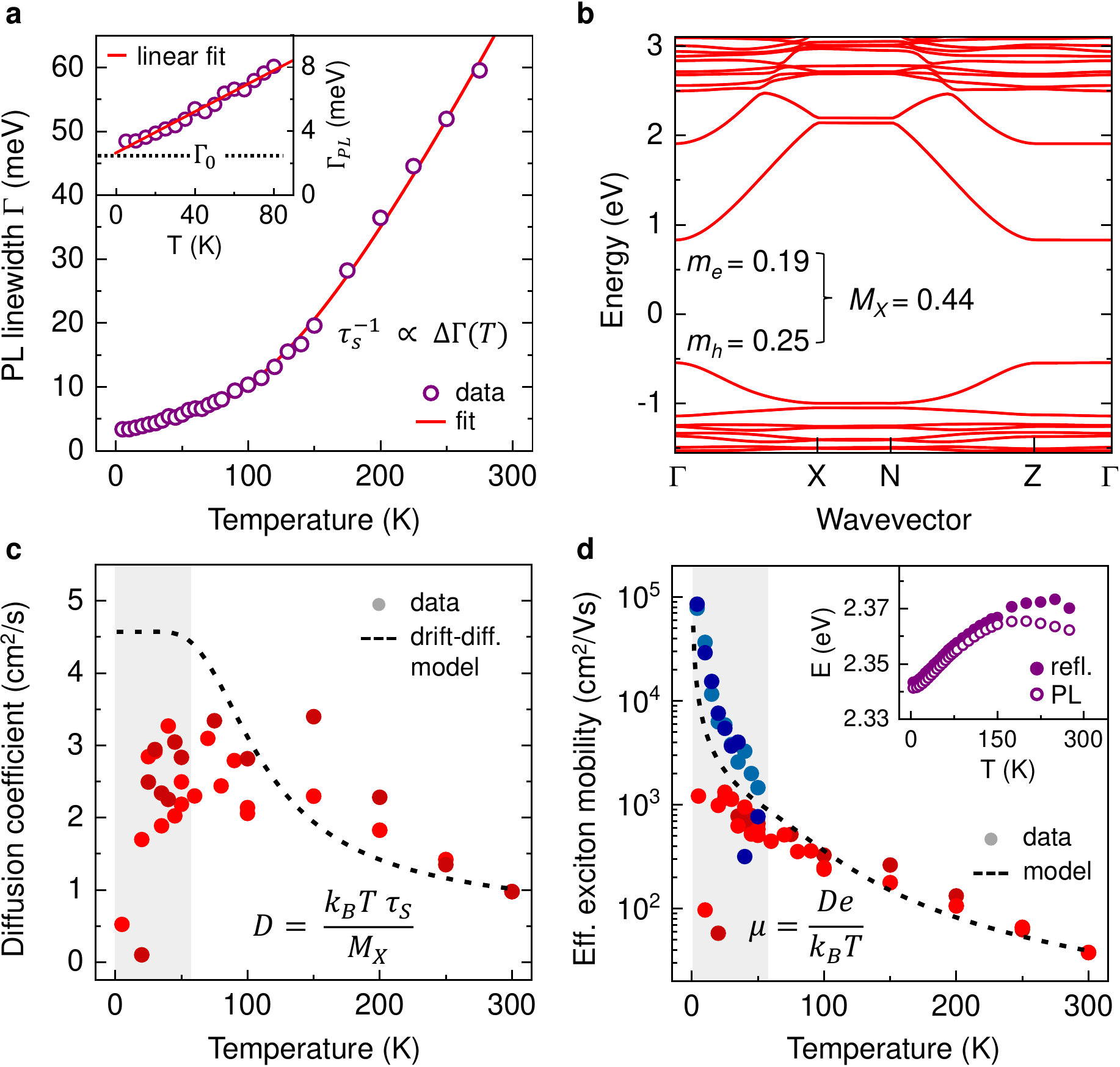}
		\caption{Exciton-phonon scattering, exciton diffusion and effective mobility.  
		a. Measured temperature-dependent linewidths from continuous-wave PL together with a phenomenological fit. 
		Linear broadening regime below 100\,K is illustrated in the inset. 
		b. Calculated electron bandstructure from first-principles used for the estimation of the effective total mass of the exciton.
		c. Estimated diffusion coefficients from a basic semi-classical drift-diffusion model.	
		d. Effective exciton mobility obtained from the Einstein relation.
		Temperature-dependent peak energies from steady-state reflectance and PL measurements illustrating finite Stokes shifts are presented in the inset.
		Corresponding experimental data from Fig.\,\ref{fig3} for long (red dots) and short (blue dots) time scales is added to c and d.
		Gray areas indicate the low-temperature regime of anomalous diffusion. 
		}
	\label{fig4}
\end{figure}

At elevated temperatures, the excitons exhibit conventional diffusion with a linear increase of the MSD over time, accompanied by comparatively long, nearly single-exponential decay of the PL on the order of several 100's of ps up to a few ns (see SI). 
Corresponding values for the effective diffusion coefficients are summarized in Fig.\,\ref{fig3}c.
Importantly, above roughly 50\,K, the diffusivity remains either constant or decreases with increasing temperature, i.e., $\partial D/\partial T \leq 0$, which is a characteristic hallmark of \textit{free propagation}\,\cite{Gantmakher1987}.
It also implies that defect-related scattering is negligible in this range, as it would lead to the increase of diffusivity with temperature otherwise.
In addition, the slope on the order -0.01\,cm$^2$/sK at higher temperatures is further consistent with the observations around room temperature reported in Ref.\,\cite{Seitz2020} 
The obtained diffusion lengths, presented in the inset of Fig.\,\ref{fig3}c, increase up to 1\,$\mu$m with the corresponding effective mobilities reaching up to 1000\,cm$^2$/Vs (also see Fig.\,\ref{fig4}d). 

In stark contrast to that, exciton propagation at lower temperatures is \emph{anomalous}, exhibiting subdiffusive dynamics below 50\,K and even regions of negative effective diffusion coefficient between 5 and 20\,K, independent from injection density (see SI).
The exciton cloud spreads rapidly during the first 100\,ps with extracted diffusion coefficients up to 30\,cm$^2$/s, as shown in Fig.\,\ref{fig3}c by blue dots, and decays within about 40$\pm10$\,ps.
Interestingly, the diffusivity of this fast component also decreases with temperature, indicating free propagation.
In addition, the combination of fast diffusion and short lifetime still results in sizable diffusion lengths on the order of many 100's of nm.
After the initial recombination, the remaining exciton population decays very slowly with the diffusion coefficients approaching zero at long time scales at 5\,K.
However, the diffusivity of this long-lived component rapidly increases with temperature (see red dots in Fig.\,\ref{fig3}c for $T<50$\,K), characteristic for a progressive thermal activation of a fraction of excitons being either localized\,\cite{Mott1969,Shklovskii1984} or scattered by imperfections\,\cite{Oberhauser1993}.
In the following section we discuss our findings and outline pathways to rationalize and interpret the observed behavior.

\textbf{Discussion of Exciton Diffusion Mechanisms.}
Based on the observation of free exciton propagation across a broad range of temperatures it is reasonable to consider a basic semi-classical drift-diffusion model that is commonly used to describe band-like transport in inorganic semiconductors\,\cite{Ashcroft1976, Oberhauser1993}.
In this model, an exciton with a total mass $M_X$ moves with an average thermal energy $k_BT$ and randomly changes the direction due to momentum scattering on a characteristic time scale $\tau_s$, yielding the diffusivity 
\begin{equation}
D=\frac{k_BT\tau_s}{M_X}.
\label{drift-diffusion}
\end{equation} 

Experimentally, $\tau_s$ is estimated from the temperature-dependent spectral broadening $\Delta\Gamma(T)$ of the exciton resonance according to $\tau_s = \hbar/\Delta\Gamma(T)$.
Total linewidth $\Gamma=\Gamma_0+\Delta\Gamma(T)$ is obtained from spectrally-resolved continuous-wave PL measurements and presented in Fig.\,\ref{fig4}a as full-width-at-half-maximum values from Lorentz peak fits.
The constant offset $\Gamma_0=3.1$\,meV is attributed to radiative and residual inhomogeneous broadening that should not contribute to the momentum scattering.
The linewidth increases strongly with temperature, characteristic for efficient thermally activated exciton-phonon scattering in 2D perovskites\,\cite{Guo2016, Wright2016}. 
It is well described in the Debye-Einstein approximation\,\cite{Wright2016} $\Delta\Gamma(T)=\gamma_AT+\Gamma_O/(exp[E_O/k_BT]-1)$, phenomenologically accounting for thermal populations of acoustic and optical modes.
Here, the parameters are $\gamma_A=50$\,$\mu$eV/K, $\Gamma_O=165$\,meV, and $E_O=37$\,meV.
We emphasize that they do not necessarily correspond to actual phonon modes of a complex vibrational structure and are used to provide effective values to describe temperature-dependent linewidth broadening.
For the estimation of the exciton effective mass, we perform first-principles calculations of the electronic bandstructure (see Fig.\,\ref{fig4}b, and SI for full details), using the VASP code\,\cite{Kresse1996}, PBE functional\,\cite{Perdew1996} including spin-orbit coupling, and TS dispersive corrections\,\cite{Tkatchenko2009}.
The effective masses of the electrons and holes, close to the fundamental band gap of the material at the $\Gamma$-point are $m_e=$\,0.19\,$m_0$ and $m_h=$\,0.24\,$m_0$ with $m_0$ denoting free electron mass.
The total mass of the exciton is thus $M_X=m_e+m_h=0.44$\,$m_0$.

The resulting temperature-dependent diffusivity obtained from Eq\,\eqref{drift-diffusion} using the above values for $\tau_s$ and $M_X$ is presented in Fig.\,\ref{fig4}c together with the experimental data from the long-lived component.
In addition, Fig.\,\ref{fig4}d shows the corresponding effective exciton mobilities from the Einstein relation, also including values from the short-lived population at low temperatures.
It is interesting to see that the simple model reasonably captures both the overall temperature dependence and the absolute values, in particular for the free-propagation regime at elevated temperatures.
It strongly implies that the observed behavior is likely to originate in efficient exciton-phonon scattering determining the diffusion.
Good quantitative agreement at room temperature further suggests that the experimentally obtained values are likely to be very close to the intrinsic limit of diffusivity, consistent with the high optical quality of the studied flakes with very small inhomogeneous broadening.

We note, however, that there are several important considerations for the interpretation of our findings beyond the scope of the basic Drude-like model presented above. 
First, there are observations of nonlinear, anomalous diffusion dynamics at low temperatures (Figs.\,\ref{fig3}a and b) that suggest an interplay of multiple components. 
Here, the emission from the rapidly diffusing and decaying excitons and a long-lived, barely mobile fraction can indeed lead to an overall shrinking of the spatial distribution of the detected signal and effectively result in apparent negative diffusivity.
Moreover, the assumption of the exciton temperature being equal to that of the lattice may not hold in this regime.
Non-resonantly injected excitons could relax on comparable or slower timescales as those for recombination, leading to an overestimation of the experimentally extracted mobility after the excitation.
A possible scenario would involve rapid diffusion of hot excitons that subsequently cool down and localize at low temperatures. 
Elevating the temperature then increasingly impedes hot exciton propagation due to faster scattering with phonons but also effectively lifts localization for the long-lived exciton fraction. 

Secondly, Eq.\,\eqref{drift-diffusion} was used under the assumption of a constant effective mass.
In case of strong exciton-phonon coupling the excitons may be better described in the polaron picture\,\cite{Gauthron2010,Guo2016,SrimathKandada2020} with a renormalized and, most importantly, temperature-dependent mass. 
Precedent for the impact of such effects can be found in a recent theoretical work on 3D perovskites demonstrating strong fluctuations in the electronic coupling that are driven by temperature-induced lattice dynamics\,\cite{Mayers2018}.
Phenomenologically, the decrease of diffusivity with temperature in Eq.\,\eqref{drift-diffusion} would then stem from the temperature dependence of both momentum scattering time and effective mass renormalization of the polarons. 
The observation of finite Stokes shifts between reflectance and emission peak energies that seem to increase with temperature, as illustrated in the inset of Fig.\,\ref{fig4}d, provides additional support for the potential relevance of these effects\,\cite{Guo2019}. 

Finally, we note that the applicability of a semi-classical model is formally limited to the mean free path being comparable or larger than the de Broglie wavelength of the propagating exciton wavepackets.
In our case, this condition does not hold at elevated temperatures where the scattering rate strongly increases even as the predictions of the model match experimental data very well in this regime.
For a more adequate microscopic description it may thus require additional quantum interference corrections\,\cite{Glazov2020} or full quantum-mechanical treatment.
We thus emphasize that the basic description presented above is intended as one of possible pathways to rationalize key experimental findings and highlight the role of the exciton-phonon coupling.
Consequently, it strongly motivates further theoretical inquiries into the physics involved.

\textbf{Conclusions.}
In summary, we have experimentally demonstrated fast exciton diffusion in 2D hybrid perovskites from room temperature down to liquid helium conditions, across conventional and anomalous diffusion regimes.
We found characteristic hallmarks of free particle propagation of long-lived exciton populations for all temperatures above 50\,K with the effective mobilities up to 1000\,cm$^2$/Vs.
At lower temperatures we observed rapid initial diffusion followed by a non-monotonous evolution of the exciton distribution and subsequent localization at longer time scales.
A drift-diffusion analysis using experimentally determined scattering rates and a theoretically computed exciton mass allowed for a reasonable description of the linear diffusivity across the studied temperature range.
It strongly suggests that the measured exciton mobilities are likely to be close to intrinsic limits determined by efficient exciton-phonon coupling.
Altogether, the reported findings render the hybrid low-dimensional perovskites as exceptionally suitable systems both for fundamental studies of strongly-interacting many-body states and optoelectronic applications involving mobile optical excitations.

\subsection*{Acknowledgements}
We thank our colleagues Mikhail Glazov, Paulina Plochocka, Sivan Refaely-Abramson, Claudia Draxl, and Henry Snaith for helpful discussions, Robin P. Puchert for the support with AFM measurements and John M. Lupton for sharing the AFM.
Financial support by the DFG via SPP2196 Priority Program (CH 1672/3-1) and Emmy Noether Initiative (CH 1672/1-1) is gratefully acknowledged. 
D.A.E. acknowledges funding from: the Alexander von Humboldt Foundation within the framework of the Sofja Kovalevskaja Award, endowed by the German Federal Ministry of Education and Research; the Technical University of Munich - Institute for Advanced Study, funded by the German Excellence Initiative and the European Union Seventh Framework Programme under Grant Agreement No. 291763; the DFG under Germany's Excellence Strategy - EXC 2089/1–390776260.
O. Y. acknowledges funding from ISF (1861/17) and ERC (850041 - ANHARMONIC). 
K.W. and T.T. acknowledge support from the Elemental Strategy Initiative conducted by the MEXT, Japan, Grant Number JPMXP0112101001, JSPS KAKENHI Grant Numbers JP20H00354 and the CREST(JPMJCR15F3), JST.




\newpage
\section{\center{Supplementary Information}}

\section{Methods}
\subsection {Crystal synthesis}
For the material synthesis we used PbO (99\%), phosphinic acid (50\%w in H$_2$O), 2-phenylethyl-amin (99\%) and n-butylamine (99.5\%), purchased from Merck, hydriodic acid (stabilizer free, 57\%w in H$_2$O), purchased from Holland Moran, and diethylether (BHT stabilized) purchased from Bio-Lab Chemicals. 
These reagents were used without any additional treatment. 
Single crystals of phenylethylammonium-lead-iodide, (C$_6$H$_5$-CH$_2$CH$_2$NH$_3$)$_2$PbI$_4$, were synthesized according to procedure reported in Refs.\,\cite{Yaffe2015,Stoumpos2016} and adapted according to Refs.\,\cite{Peng2017,Chen2018}. 
In brief, PbO (0.3 g, 1.34 mmol) was dissolved in a mixture of hydriodic acid (1.3\,mL, 10.2\,mmol) and phosphinic acid (228\,$\,u$L, 2\,mmol) at 110$^o$C under magnetic stirring to a clear yellow solution. 
In a separated beaker held in an ice-bath, 2-phenethylamine solution (171\,$\mu$L, 1.34\,mmol) was neutralized with hydriodic acid (11\,mL, 83.5\,mmol) and added to the PbI$_2$ solution, immediately leading to some precipitation. 
The magnetic stirring was continued until the solution was again of clear yellow color. 
Hydriodic acid was added drop wise (about 1\,mL) to speed up the dissolution. 
The stirring was stopped and the beaker was capped and cooled to room temperature at a rate of 2\,K/hour. 
The orange plate-like crystals were further quenched in an ice bath, vacuum filtered, washed with diethylether, and dried at 90$^o$C in a vacuum oven for 24 hours prior to use.
The material composition and crystal structure were confirmed by x-ray diffraction and Raman spectroscopy.

\subsection {Preparation of encapsulated samples}
To obtain hBN-encapsulated samples we used the technique outlined in Ref.\,\cite{Castellanos-Gomez2014} for inorganic van der Waals materials and applied in Ref.\,\cite{Seitz2019} for 2D perovskites.
Thin hBN flakes were first mechanically exfoliated on polydimethylsiloxane (PDMS) from bulk crystals (provided by T. Taniguchi and K. Watanabe, NIMS) and subsequently stamped on a 90\,$^\circ$C preheated SiO$_2$/Si substrate at ambient conditions.
Thin 2D perovskite flakes, also exfoliated on PDMS, were then stamped on top of the hBN layers followed by the stamping of a second thin hBN layer on top.
Particular care was taken to provide full contact of the bottom and top hBN layers outside the perovskite sample to ensure complete encapsulation.
We also note that the thickness of the perovskites sheets was kept roughly below 100\,nm for optimal heterostructure fabrication.
Layer thicknesses were estimated from optical contrast and confirmed by atomic-force microscopy.

\subsection {Optical spectroscopy}
For time-resolved measurements we used a Ti:sapphire laser source (Chameleon Ultra II, Coherent Inc.) for excitation, providing 100\,fs-short pulses with a repetition rate of 80\,MHz and photon energies of 3.1\,eV obtained through second-harmonic generation of the fundamental beam.
The laser was focused on the sample held in an optical microscopy cryostat (ST500, Janis) by a 60$\times$ glass-corrected microscope objective (Nikon) on a spot size of about 1\,$\mu$m in diameter.
Excitation densities were varied between 0.3 and 130\,nJ/cm$^2$.
Assuming roughly $10\%$ absorption per layer\,\cite{Yaffe2015} at the excitation photon energy, these values would correspond to the maximum injected electron-hole densities between $10^8$ and 3$\times10^{10}$\,1/cm$^2$ per layer. 
For continuous-wave PL and reflectance measurements we used a 473\,nm laser source (excitation power density of 0.25\,W/cm$^2$) and a tungsten-halogen lamp, respectively.
The sample emission was imaged onto a spectrometer slit and subsequently onto a streak camera detector (C10910-05, Hamamatsu), using either a grating (300\,gr/cm or 1200\,gr/cm) or a silver mirror for spectrally- and spatially-resolved measurements, respectively.
The streak camera was operated in single-photon counting mode and provided time-resolution of the luminescence signal on the order of several ps to a few 10's of ps depending on the chosen temporal range.
Further details on the spatially- and time-resolved microscopy technique, measurement procedure, and data analysis are found in Ref.\,\cite{Kulig2018}.
For time-integrated detection a peltier-cooled CCD (Pixis256, Roper Scientific) was employed.

\subsection {Bandstructure calculations} 
Density functional theory calculations were performed using the ``Vienna Ab initio Simulation Package'' code\,\cite{Kresse1996}, projector augmented wave potentials\,\cite{Kresse1999a} and accounting for spin-orbit coupling (SOC). 
Exchange-correlation was treated using the Perdew-Burke-Ernzerhof (PBE) functional\,\cite{Perdew1996} augmented by dispersive corrections that were computed using the Tkatchenko-Scheffler method\,\cite{Tkatchenko2009}.
The plane-wave kinetic energy cutoff was set to 500\,eV and a 4$\times$4$\times$1 $\Gamma$-centered k-point grid was used for self-consistently calculating the charge density. 
The experimentally-determined crystal structure of ($n=1$) PEA$_2$PbI$_4$ was taken as a starting point to optimize the lattice parameters and atomic positions, in calculations without accounting for SOC, until the force on each atom was smaller than 0.01\,eV\AA$^{-1}$.
In this way, we obtained a unit cell with lattice constants of $a=8.48$\,\AA, $b=8.89$\,\AA, and $c=16.40$\,\AA, which agreed well with the experimental result of $a=8.67 $\,\AA, $b = 8.68$\,\AA, and $c = 16.40$\,\AA~from x-ray diffraction. 
The differences in the in-plane lattice constants was found to be due to the orientation of the PEA molecules, which can be improved using a supercell that allows for a more realistic representation of their geometry. 
However, we verified this to have no effect on the effective mass results. 
The electronic bandstructure calculations were performed non-self-consistently, using an equally-spaced k-grid of 50 points on each path between two high symmetry points ($\Delta k\approx 0.001$\,\AA$^{-1}$); the numerical convergence with respect to $\Delta k$ was verified.
Electron and hole masses, $m_e^*$ and $m_h^*$ were calculated around $\Gamma$ towards the X-point (0.5, 0, 0) in the Brillouin zone, since the fundamental gap occurs at the $\Gamma$-point. 
The effective masses were obtained using a finite difference method following the equation $m^*=\hbar^2(\partial^2 E/\partial^2 k)^{-1}\approx\hbar^2((E(k_\Gamma+2\Delta k)+E(k_\Gamma)-2E(k_\Gamma+\Delta k))/\Delta k^2)^{-1}$.
The exciton total effective mass $M_X$ was then obtained according to $M_X = m_e^* + m_h^*$.
To test the reliability of our PBE + SOC results, we have also performed calculations using the HSE functional\,\cite{Heyd2003,Heyd2006}, from which we obtained an identical result for the reduced exciton mass compared to the PBE data.

\newpage

\section{Height profiles from atomic-force microscopy}

\begin{figure}[h]
	\centering
		\includegraphics[width=17 cm]{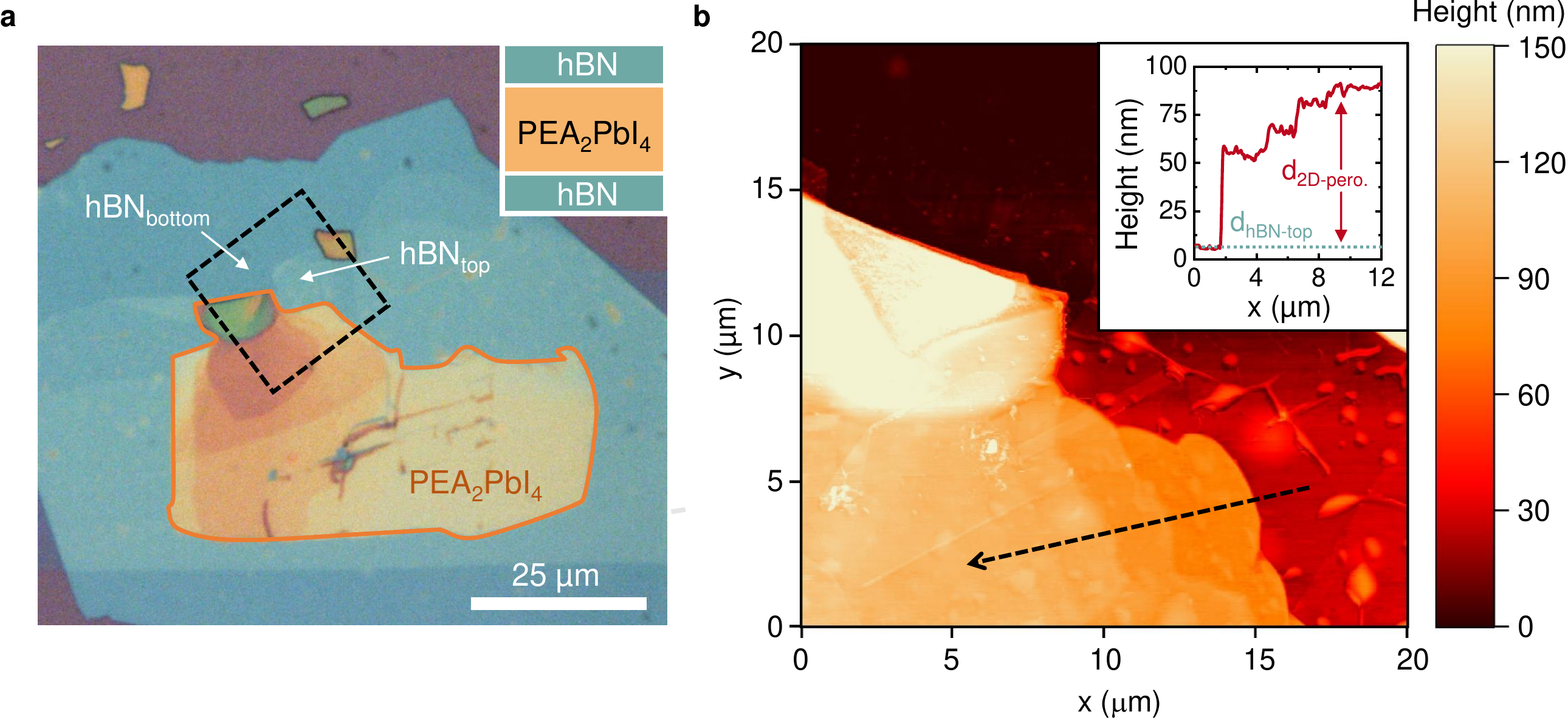}
		\caption{\textbf{Characterization of hBN-encapsulated PEA$_2$PbI$_4$.} 
		\textbf{a.} Optical micrograph of a typical PEA$_2$PbI$_4$ perovskite flake encapsulated between two thin layers of hexagonal boron nitride. 
		The black rectangle shows the area scanned with the AFM. 
		\textbf{b.} Corresponding AFM image of the topography.
		A cross-section along the black dashed line is presented in the inset.
		Vertical offset at $x=0$\,$\mu$m corresponds to the thickness of the top hBN layer followed by a step-like increase of the thickness of the underlying perovskite flake.}
\label{figSI_1}
\end{figure}

To determine the layer thickness and confirm the values obtained from the analysis of reflectance spectra, we performed atomic-force microscopy (AFM) measurements on one of the studied hBN-encapsulated 2D perovskite (PEA$_2$PbI$_4$) flakes.
For this purpose, we chose a sample with a particularly thin top layer hBN (as estimated from optical contrast) that is expected to closely follow the underlying, much thicker perovskite layer.
The scanned region, indicated in the optical micrograph in \fig{figSI_1}(a), was selected to include both the edge of the hBN capping layer on top of the bottom hBN as well as a part of the encapsulated structure with the perovskite layer sandwiched in-between.
AFM images were taken with silicon probes (Nanosensors PPP-NCHR and SSS-NCHR) using a Park Systems XE-100 operating in ‘non-contact’ amplitude modulation mode under ambient conditions in a vibration-isolated acoustic enclosure.

The AFM topography image is presented in \fig{figSI_1}(b) and a selected trace across the perovskite layer is shown in the inset.
The vertical offset of about 8\,nm at $x=0$\,$\mu$m corresponds to the top hBN layer thickness obtained by measuring an edge where both encapsulating hBN layers are in direct contact.
This parameter is also used for subsequent analysis of the optical response in the presence of multilayer interference effects, as described in more detail further below. 
In addition, since hBN usually exhibits atomically flat surfaces and the top hBN layer is already very thin, the measured surface profile should largely follow the height of the perovskite flake underneath. 
The obtained thickness varies between 50 to 80\,nm in step-like changes that are typical for exfoliated, layered van der Waals materials.
Importantly, due to the high optical contrast of the perovskite layers these steps are also visible in the optical micrograph.
In spatially- and time-resolved emission experiments, the measurement spots with a diameter of about 1\,$\mu$m were chosen to be well within the plateau-like region of 80\,nm thickness.

\newpage
\section{Stability under pulsed excitation}

\begin{figure}[h]
	\centering
		\includegraphics[width=16 cm]{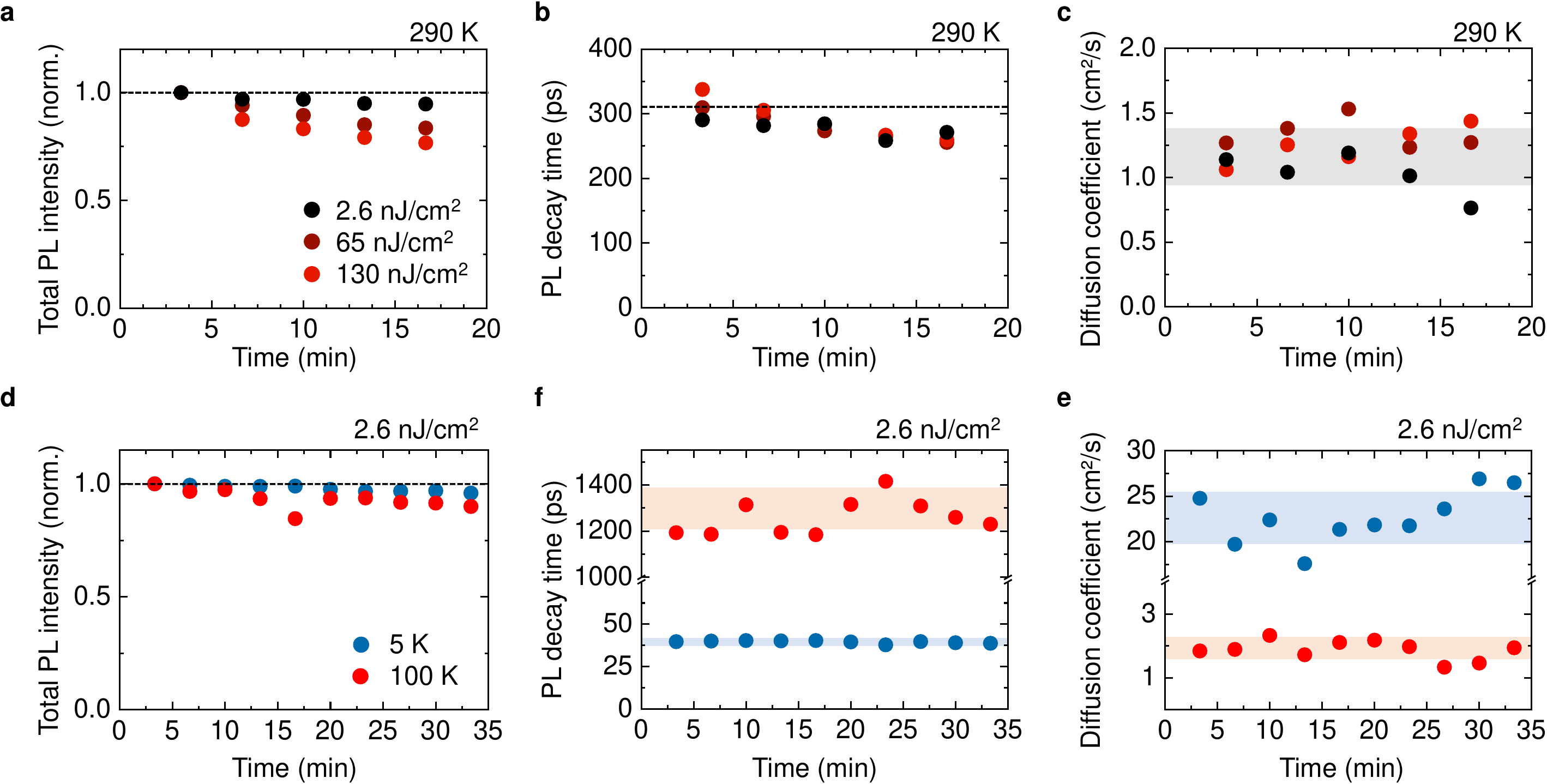}
		\caption{\textbf{Sample stability during measurements under pulsed optical excitation.} 
		\textbf{a.} Normalized total PL intensity as function of measurement time for different excitation densities at room temperature. 
		\textbf{b.} PL decay time from single-exponential fitting of the transients. 
		\textbf{c.} Corresponding diffusivity from spatially-resolved emission.
		Labels in \textbf{b} and \textbf{c} are the same as in \textbf{a}.
		\textbf{d.} Integrated PL normalized to the initial PL intensity as function of measurement time for 5\,K and 100\,K at a low excitation power of 2.6\,nJ/cm$^2$ per pulse. 
		Corresponding PL decay time and diffusivity are presented in \textbf{e} and \textbf{f}, respectively for 5\,K (blue circles) and 100\,K (red circles).}
\label{figSI_2}
\end{figure}
Exciton diffusion measurements in our study were performed on perovskite layers encapsulated in hBN and studied under high vacuum conditions to reduce degradation effects by effectively protecting the material from the environment \cite{Boyd.2019}\cite{Seitz.2019}\cite{Holler.2019}. 
\fig{figSI_2}(a) shows total PL intensity as function of measurement time for several excitation densities up to 130\,nJ/cm$^2$ per pulse, where the laser induced degradation is expected to be most pronounced. 
Even for the highest density applied we observe only a small decrease of the total PL during a measurement cycle of more than 15 minutes.
It is accompanied by a similarly small decrease in the PL lifetime, as illustrated in \fig{figSI_2}(b).
For a comparatively low excitation density of 2.6\,nJ/cm$^2$, however, the decrease in both emission intensity and decay time is nearly absent, being as small as about 5\%. 
More importantly, the diffusion coefficient presented in \fig{figSI_2}(c), extracted from single frames recorded by the streak camera, does not show any systematic dependence on the measurement time.
That implies that even a weak decrease of the PL intensity and lifetime at the highest density, attributed to faster non-radiative recombination from degradation, does not influence the obtained diffusivity. 
Nevertheless, due to essentially absent degradation effects and still reasonable signal-to-noise ratios, a very low excitation density of 2.6\,nJ/cm$^2$ was chosen for subsequent temperature-dependent measurements. 
At these conditions, also at 5 and 100\,K, we observe only minor changes in the PL intensity after more than 30 minutes of continuous irradiation with a pulsed laser, as illustrated in \fig{figSI_2}(d).
No consistent changes are detected in the decay time (\fig{figSI_2}(e)) and diffusivity (\fig{figSI_2}(f)).

\newpage
\section{Influence of diffusion on measured emission lifetime}

\begin{figure}[h]
	\centering
		\includegraphics[width=13 cm]{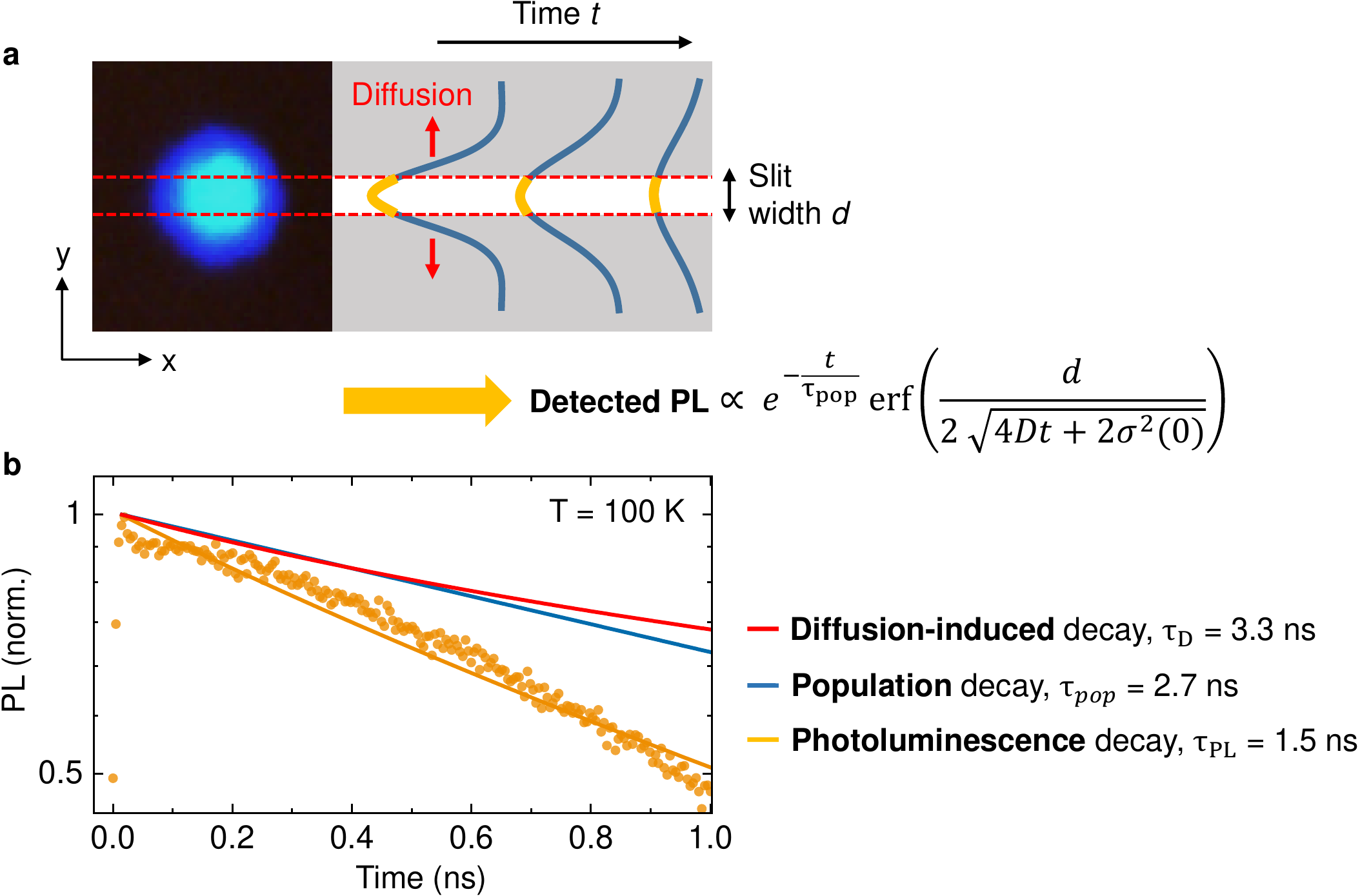}
		\caption{\textbf{Exciton lifetime analysis and correction for diffusion.} 		
	\textbf{a.} Schematic illustration of the light emission from diffusing excitons in the employed, effectively one-dimensional geometry.
	The excitons diffuse in two dimensions and thus move out of the detection area leading to an additional decay channel in the measured PL.
	The resulting emission is then described by the product of the population decay (e.g., exponential dependence on time with a constant $\tau_{pop}$) and an error function ($erf$) that depends on the diffusivity $D$ and the spatial variance $\sigma^2(0)$ immediately after the excitation, i.e., at $t$=0.
	\textbf{b.} Exemplary fit at 100\,K, where the influence of the diffusion is very pronounced. 
	The PL loss due to diffusive processes occurs on a time scale of $\tau_{D}$=3.3 ns, leading to a corrected exciton lifetime $\tau_{pop}$=2.7 ns compared to the measured PL lifetime of $\tau_{PL}$=1.5 ns.}
\label{figSI_3}
\end{figure}

Here, we discuss the influence of the exciton diffusion on the measured PL lifetime, as schematically illustrated in \fig{figSI_3}(a).
While the excitons emit and diffuse in two dimensions, the detected signal is collected from an effectively one-dimensional cross-section using a fixed slit width $d$ in the imaging plane. 
In addition to the population decay of the excitons associated with actual recombination, the excitons thus also diffuse out of the detection area (along the vertical coordinate in the illustration in \fig{figSI_3}(a)).
This diffusion gives rise to an additional decay channel in the measured PL intensity that we need to take into account to extract the actual dynamics of the exciton population.
The fraction of excitons in the detection cross-section is obtained from the area of a Gaussian profile from $-d/2$ to $+d/2$ with the time-dependent standard deviation of $\sqrt{2Dt+\sigma^2(0)}$, that increases with time due to diffusion.
The resulting PL transient is then described by a product of time-dependent population (e.g., exponentially decaying with a constant $\tau_{pop}$) and an error function $erf(d/2\sqrt{4Dt+2\sigma^2(0)})$ that depends on the diffusivity $D$ and the spatial variance of the spot profile $\sigma^2(0)$ immediately after the excitation.

A representative case is shown in \fig{figSI_3}(b) using the measured data at 100\,K, where the influence of diffusion on the emission dynamics is particularly strong.
The PL decays with a characteristic time-constant $\tau_{PL}$ of 1.5\,ns.
The extracted lifetime of the exciton population, however, is much longer with $\tau_{pop}=2.7$\,ns.
It is comparable to the decay constant of the diffusion-induced component, estimated by $\tau_{D} = (1/\tau_{PL} - 1/\tau_{pop})^{-1}$ = 3.3\,ns.
The above procedure is then consistently applied to extract exciton lifetimes in the presence of diffusion.
We note, however, that the influence of the latter is less pronounced both at higher and lower temperatures than 100\,K due to lower diffusion coefficients and shorter population lifetimes, respectively.

\newpage
\section{Density- and temperature-dependent decay dynamics}

\begin{figure}[h]
	\centering
		\includegraphics[ width=16 cm]{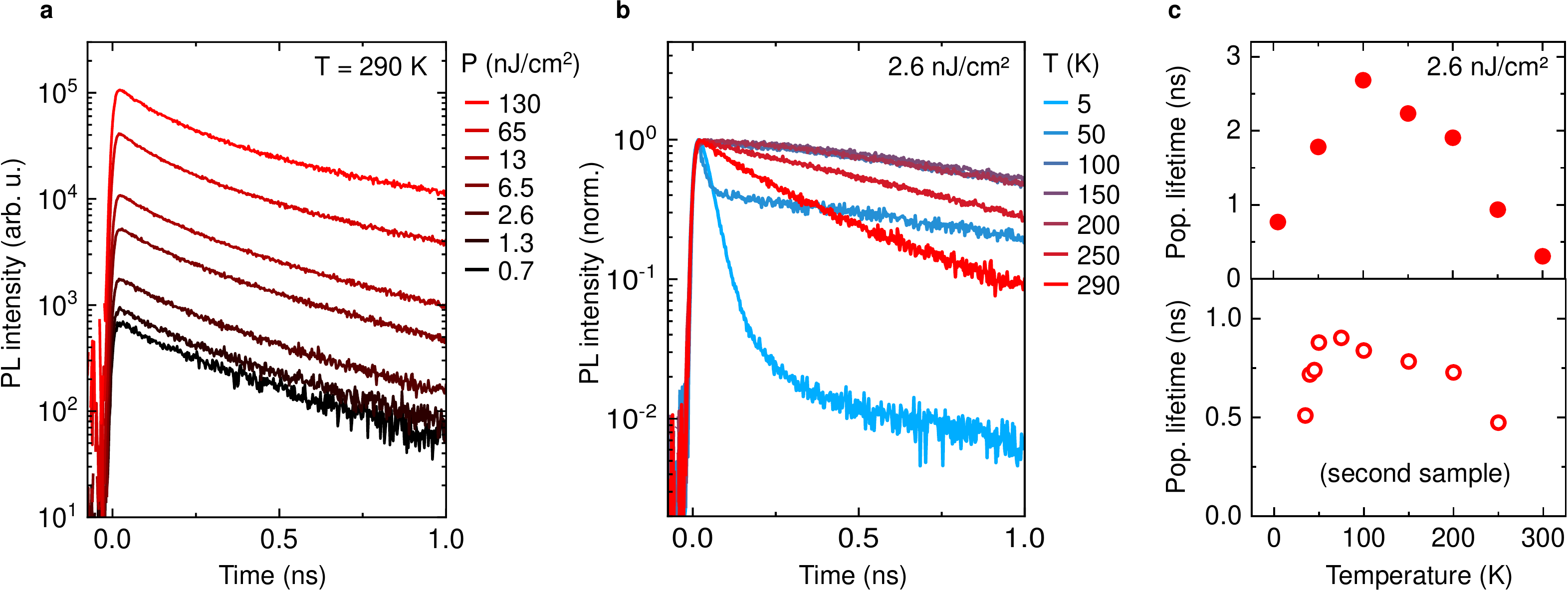}
		\caption{\textbf{Photoluminescence dynamics at different excitation powers and temperatures.} 
		\textbf{a.} As-measured room temperature PL transients for different excitation densities.  
		\textbf{b.} Normalized PL transients measured in the lower excitation regime at 2.6 nJ/cm$^2$ at temperatures from 5 to 290\,K. 
		\textbf{c.} (upper panel): Extracted exciton lifetimes from the long-lived component.
		(lower panel): Exciton lifetimes obtained from a second sample.
		}
\label{figSI_4}
\end{figure}

\fig{figSI_4}(a) shows PL transients at $T=290$\,K spatially-integrated over the measured cross-section for several excitation densities from 0.7 to 130\,nJ/cm$^2$.
All traces are very similar and decay nearly exponentially on a time scale of about 0.3\,ns.
Normalized PL transients for varying temperatures from 290\,K down to 5\,K are presented in \fig{figSI_4}(b).
For all temperatures down to 50\,K they exhibit a mono-exponential decay with the extracted population lifetime increasing from 0.3\,ns at room temperature to almost 3\,ns at 100\,K, as illustrated in the upper panel of \fig{figSI_4}(c).
We note, however, that we observed variations in the absolute values of the exciton lifetime for different studied flakes at lower temperatures, as shown in the lower panel of \fig{figSI_4}(c).
Importantly, despite these differences in the exciton lifetime, the diffusivity values and their temperature dependence are very similar for the two samples, as presented in the main manuscript.
 
\cleardoublepage

\section{Exciton diffusion and decay at low temperatures}
\label{section4}
\begin{figure}[h]
	\centering
		\includegraphics[width=12 cm]{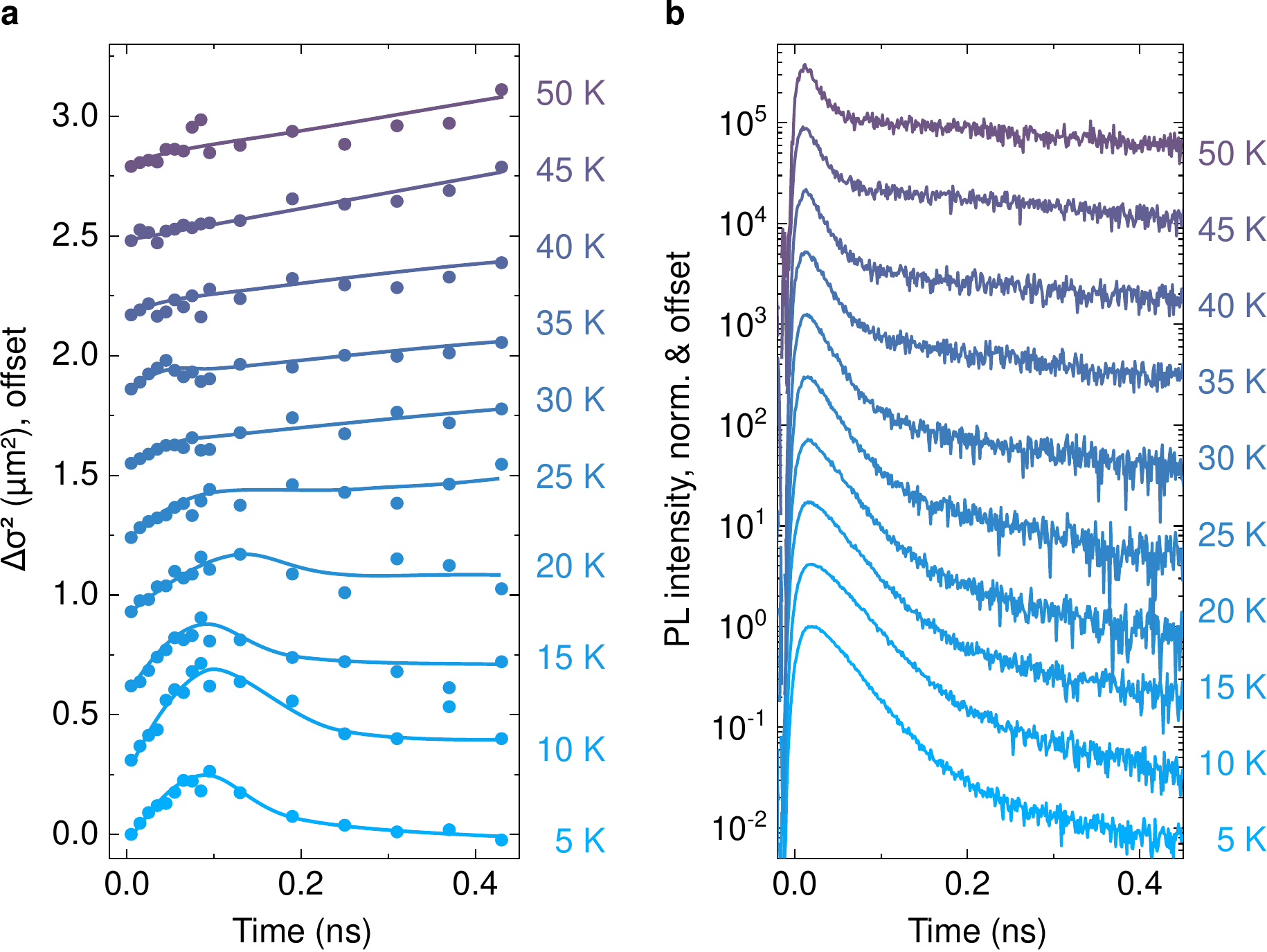}
		\caption{\textbf{Mean squared displacement and transient PL at low temperatures}
		\textbf{a.} Time-resolved mean squared displacement $\Delta \sigma^2 =  \sigma^2$(t) -  $\sigma^2$(0) of the PL profiles in the low temperature regime, from 5 to 50\,K. 
		Solid lines are guides to the eye.
		\textbf{b.} Corresponding PL transients in the same temperature range illustrating a decreasing contribution of the fast decay fraction with rising temperature.
}
\label{figSI_5}
\end{figure}

\begin{figure}[h]
	\centering
		\includegraphics[width=9.5 cm]{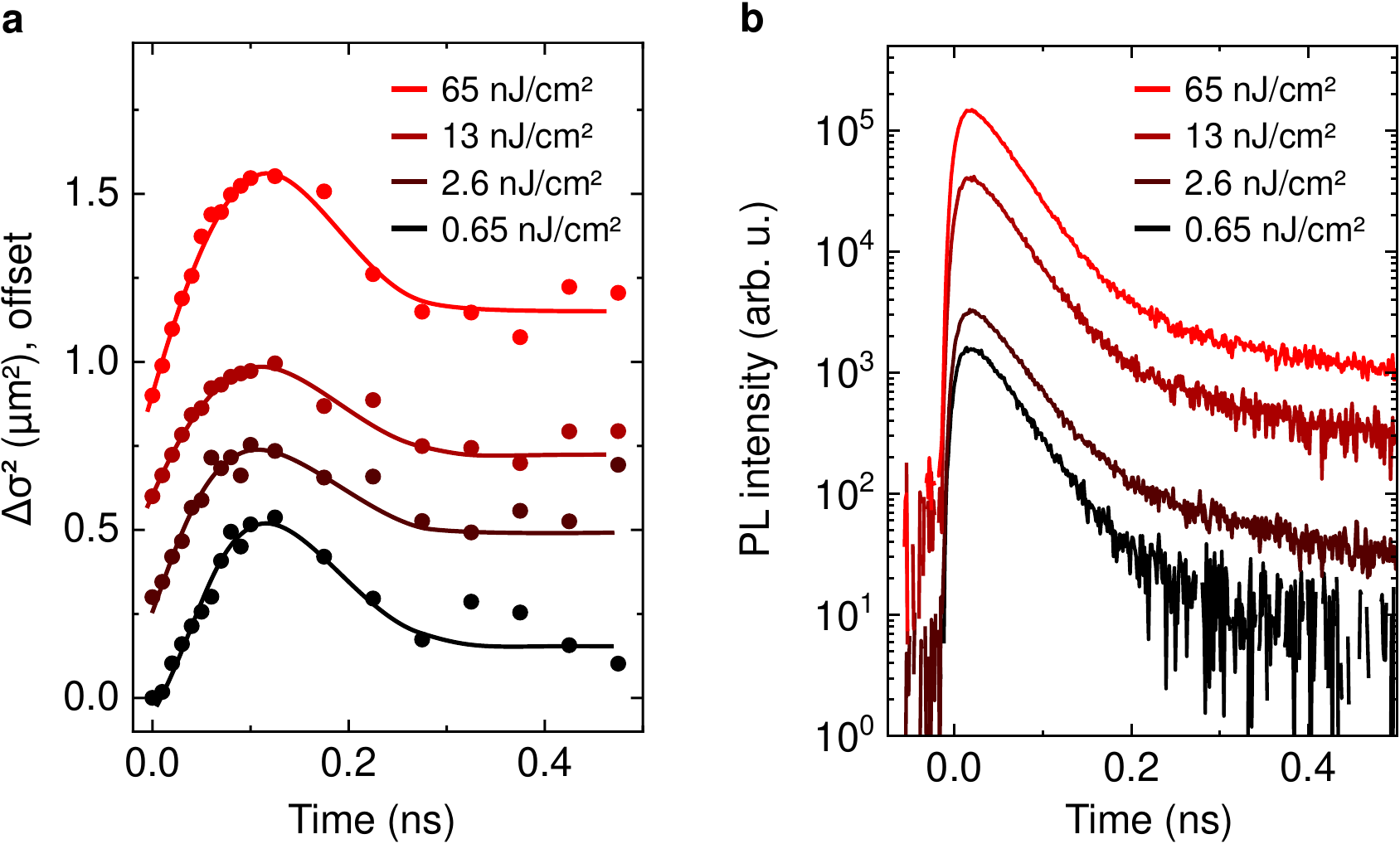}
		\caption{\textbf{Influence of excitation density at 5\,K.}
		\textbf{a.} Time-resolved mean square displacement at 5\,K of the emission across two orders of magnitude of energy density per pulse. 
		Solid lines are guides to the eye. 
		\textbf{b.} Corresponding PL transients.
		}
\label{figSI_6}
\end{figure}

\fig{figSI_5}(a) presents the time-resolved mean squared displacement (MSD) $\Delta \sigma^2 =  \sigma^2$(t) -  $\sigma^2$(0) for low temperatures between 5 and 50\,K. 
The MSD exhibits a linear increase with time at 50\,K and an anomalous behavior at 5\,K, with an initially rapid diffusion followed by an apparent shrinking of the spatial profile.
The latter is accompanied by an increasing decay component with a very short lifetime of the exciton population, as illustrated in \fig{figSI_5}(b) by the PL transients.
The two regimes merge smoothly in the presented temperature range.
In particular, negative diffusivity is observed from 5 up to 20\,K and the non-linear behavior with an initially fast diffusion persists up to 30-35\,K.
In addition, both diffusion and decay dynamics do not exhibit any pronounced excitation density dependence in the studied range at 5\,K, as illustrated in \fig{figSI_6}(a) and (b), respectively.
The underlying processes thus seem to be driven by thermal phenomena such as, e.g., changes in the phonon population, decreasing kinetic energy or relaxation rate of the excitons, rather than exciton-exciton interactions. 

\newpage
\cleardoublepage

\section{Spatial dynamics of spectrally filtered emission at 5\,K}
\label{section5}
\begin{figure}[h]
	\centering
		\includegraphics[width=13 cm]{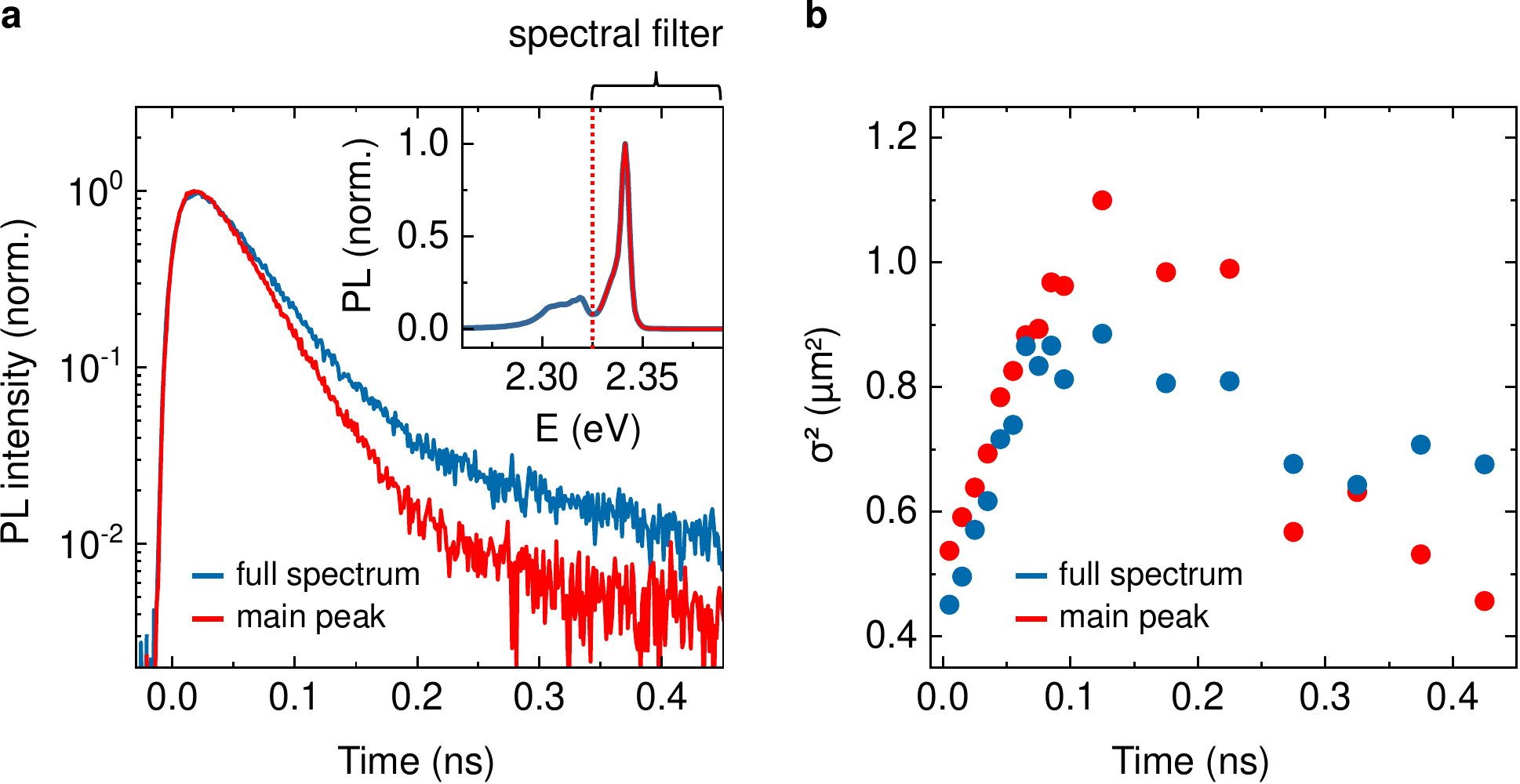}
		\caption{\textbf{Time- and spatially-resolved PL from spectrally-filtered emission at 5\,K.}
		\textbf{a.} Normalized PL transients obtained with (``main peak'', red) and without (``full spectrum'', blue) a tunable bandpass filter. 
		The inset shows the PL spectrum at 5\,K with the dotted line indicating the spectral cut-off.		
		\textbf{b.} Time-dependent variance from the spatial profiles corresponding to two measurements with (red) and without (blue) the bandpass filter.
		}
\label{figSI_7}
\end{figure}

Spatially-resolved emission reported in the main manuscript was obtained by detecting the full spectral range of the luminescence that is dominated by a single exciton peak at all temperatures (also see the following section). 
Only at very low temperatures we observed weak additional features on the low-energy side of the spectra.
To monitor their influence on the measured time- and spatially-resolved response, we used a spectrally tunable bandpass filter to compare the signal obtained from a full spectral range to that from the main exciton resonance, excluding additional emission at lower energies.
Corresponding PL transients and time-dependent variance of the spatial profile are presented in Figs.\,\ref{figSI_7}(a) and (b), respectively. 
The data in the inset of \fig{figSI_7}(a) illustrates the spectral cut-off of the bandpass filter and the resulting detection range.

Altogether, we find very similar decay and propagation dynamics in both cases. 
They are thus dominated by the main exciton resonance, consistent with comparatively weak signals from low-lying features.
The latter mainly contributes to the long-lived component of the PL that exhibits very slow diffusivity at low temperatures, approaching zero at 5\,K.
This observation also appears to be consistent with the obtained spatial variance traces in \fig{figSI_7}(b), especially if the effective shrinking of the spatial emission profile is indeed a consequence of an essentially immobile population that remains after the fast initial decay of a rapidly diffusing fraction.
Combined with the disappearance of the low-energy emission features at higher temperatures (\fig{figSI_8}), these findings indicate that these PL peaks are thus likely to stem from localized states.

\newpage
\cleardoublepage

\section{Temperature-dependent linear reflectance and emission spectra}
\label{sectionSIX}

\begin{figure}[h]
	\centering
		\includegraphics[width=15 cm]{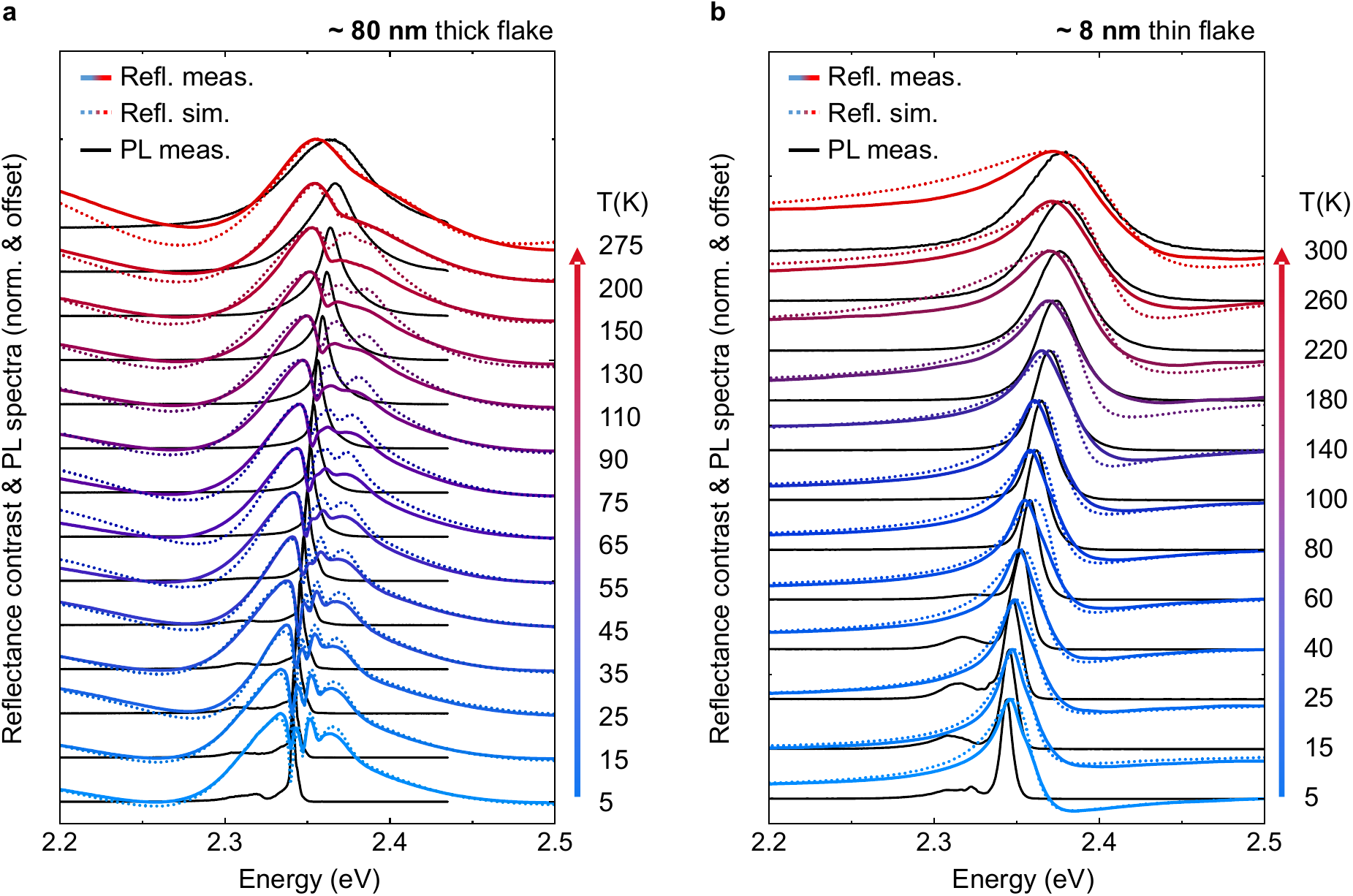}
		\caption{\textbf{Linear reflectance and PL spectra.}
		\textbf{a.} Reflectance contrast (vs. SiO$_2$/Si substrate reference) and luminescence spectra from cryogenic to room temperature measured on a thicker sample that was used for spatially-resolved experiments. 
		The data are normalized and vertically offset for clarity.
		Included are simulated reflectance contrast spectra that are obtained using multi-Lorentzian parametrization of the dielectric function and transfer-matrix method to account for multi-layer interference effects.
	  \textbf{b.} Corresponding data obtained on a thin sample with both hBN layers on top of the fused silica substrate used as a reference.
		}
\label{figSI_8}
\end{figure}

Spectrally-resolved linear reflectance and PL measured on two different hBN-encapsulated 2D perovskite (PEA$_2$PbI$_4$) samples from 5\,K to room temperature is presented in Figs.\,\ref{figSI_8}(a) and (b), respectively.
The thicker sample (\fig{figSI_8}(a)) was also used for diffusion measurements.
In addition, a much thinner flake of about 8\,nm height (\fig{figSI_8}(b)) was exfoliated and placed on fused silica substrate in order to reduce multi-layer interference effects in the reflectance.
An additional advantage of the thinner flake is that the emission should stem from the entire sample volume within the excitation spot area, while in the thicker sample we expect only the top layers to be excited due to a finite penetration depth of the pump light.

The reflectance measurements were performed both on the areas with and without the perovskite flakes, denoted as sample ($R_s$) and reference ($R_{ref}$) signals, respectively.
For the thicker sample only the underlying substrate (SiO$_2$/Si) was used as reference and for the thinner sample the reference was obtained on an area with both hBN layers on top of the substrate (fused silica).
The data is presented and analyzed as reflectance contrast that corresponds to the relative change of the reflectance in the presence of the sample.
It is defined according to $(R_s-R_{ref})/(R_{ref}-R_{bg})$, including the background counts $R_{bg}$ collected without illumination.

Reflectance contrast spectra are then analyzed by fitting the measured response with a simulated one that is obtained using a transfer-matrix approach to account for the multi-layer interference effects\,\cite{Byrnes.2016}. 
For this purpose, dielectric constants of the hBN and substrate layers are taken from literature and individual layer thicknesses are estimated from the AFM measurements. 
Frequency-dependent dielectric function of the perovskite layer in the relevant range of the main optical transition is parametrized by Lorentzian peak functions to account for exciton resonances according to:
\begin{equation} 
\varepsilon (\omega) = \varepsilon_{_{bg}} + \sum _j \frac{f_j}{E_j^2 - {E}^2(\omega) - iE (\omega) \Gamma _{j}}.
\end{equation}
Here, $f$ is the oscillator strength, $E$ is the resonance energy, $\Gamma$ is the non-radiative broadening, and $\varepsilon_{_{bg}}$ is a constant offset in the real part of the dielectric function accounting for \textit{higher-lying} resonances (it is generally not the same as the static dielectric constant).
The index $j$ then runs through exciton resonances that we include to match the resulting spectra.
The response can either be reasonably approximated with only one transition (as it is the case in \fig{figSI_8}(b)) or include several peaks (as in \fig{figSI_8}(a)).
These parameters (as well as the individual layer thicknesses to a small degree) are varied in an iterative process to yield a simulated spectrum that reasonably matches the experimentally measured response.
We further assume normal incidence conditions both for simplicity and due to a comparatively low numerical aperture of the imaging spectrometer in our experiments.
Finally, we note that while the dielectric function is defined as a three-dimensional quantity, we are effectively sensitive to the in-plane value due to the measurement geometry, where the polarization of incident and reflected light is predominantly aligned along the planes of the studied 2D perovskites.

\begin{figure}[t]
	\centering
		\includegraphics[width=16 cm]{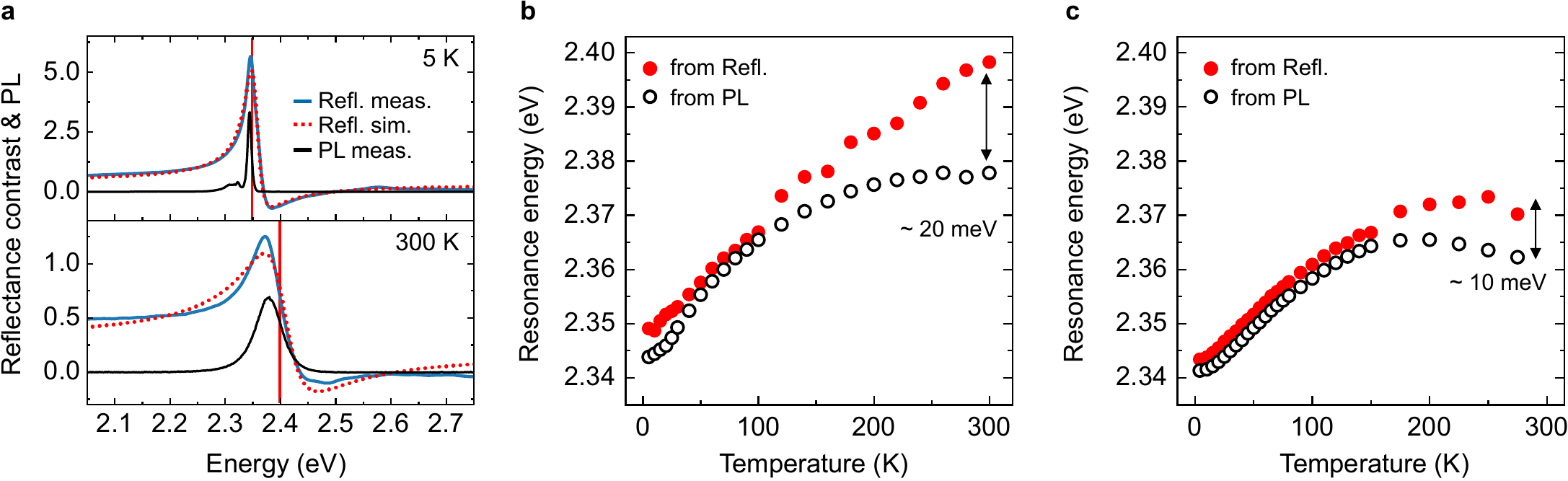}
		\caption{\textbf{Reflectance contrast and PL analysis.}
		\textbf{a.} Measured and simulated reflectance contrast spectra at 5 and 300\,K of a thin flake. 
		Resonance energies at 2.349\,eV and 2.398\,eV for 5 and 300\,K, respectively, are indicated by vertical lines.		
		\textbf{c.} Resonance energies extracted from reflectance (corresponding to those in the dielectric function) and PL as function of temperature.
		\textbf{b.} Corresponding data from a thicker flake that was used for diffusion measurements. 
		}
\label{figSI_9}
\end{figure}

The resulting simulated spectra are included in \fig{figSI_8} as dotted lines.
For comparison, measured and simulated reflectance contrast spectra from a thin sample are also presented in \fig{figSI_9}(a) together with the PL data at 5 and 300\,K.
Dashed lines indicate central energies of the exciton resonance corresponding to that of the Lorentzian peak in the dielectric function that yields simulated reflectance spectra.
Notably, these resonance energies are offset from the respective reflectance peak maxima to higher energies.
The PL spectra are fitted using symmetrical peak profiles, either Lorentzian or Voigt, that both provide very similar results.
The extracted resonance energies from reflectance, i.e., the one used in the underlying dielectric function, and PL are presented in \fig{figSI_9}(b) and \fig{figSI_9}(c) as function of temperature for the thinner and thicker samples, respectively. 
From reflectance analysis of the thicker sample (\fig{figSI_8}(a) and \fig{figSI_9}(c)), we consider the energy of the lowest energy resonance in the dielectric function.
Overall, both samples exhibit a very similar temperature dependence.
In particular, the resonance energy from the absorption-type reflectance data increases with increasing temperature that is typical for the studied materials.
At higher temperatures, however, the PL increasingly develops a finite Stokes shift with its energy being lower than that extracted from reflectance.
In three-dimensional (3D) perovskites, a Stokes shift that increases with temperature was recently attributed to polaron-type effects\,\cite{Guo2019}, albeit its magnitude was much larger, about factor of three, in contrast to our observations in 2D perovskites.



\providecommand{\latin}[1]{#1}
\makeatletter
\providecommand{\doi}
  {\begingroup\let\do\@makeother\dospecials
  \catcode`\{=1 \catcode`\}=2 \doi@aux}
\providecommand{\doi@aux}[1]{\endgroup\texttt{#1}}
\makeatother
\providecommand*\mcitethebibliography{\thebibliography}
\csname @ifundefined\endcsname{endmcitethebibliography}
  {\let\endmcitethebibliography\endthebibliography}{}

\end{document}